\documentstyle{elsart}

\def\to{\rightarrow}
\def\as{{\alpha_s}}
\def\ep{{\varepsilon}}
\def\NLO{next-to-leading order }
\def\eb{{\bar e}}
\def\qb{{\bar q}}
\def\Qb{{\bar Q}}
\def\Ps56{{\cal P}_Z(s_{56})}

\newcommand{\Nc}{{N_C}}
\newcommand{\NA}{{N_A}}
\newcommand{\Nf}{{N_f}}

\def\A60{{\cal A}^{{\rm tree}}_6}
\def\At{{\tilde A}}
\def\gt{{\tilde g}}

\def\cq{{\cal Q}}
\def\d{{\rm d}}
\def\i{{\rm i}}
\def\Tr{{\rm Tr}}
\def\bm#1{{\mbox{\boldmath $#1$}}}

\def\del#1{\lower.25em\hbox{\LARGE $\times$}\kern -1em #1 }

\newcommand{\beq}{\begin{equation}}
\newcommand{\eeq}{\end{equation}}
\newcommand{\beqn}{\begin{eqnarray}}
\newcommand{\eeqn}{\end{eqnarray}}

\def\nn{\nonumber}

\begin{document}
\begin{frontmatter}

\begin{flushright}
hep-ph/9708342
\end{flushright}

\title{Group independent color decomposition of next-to-leading order
matrix elements for $\bm e^+\bm e^-\to$ four partons}
\author[KLTE]{Zolt\'an Nagy} and
\author[ATOMKI,KLTE]{Zolt\'an Tr\'ocs\'anyi}
\address[KLTE]{Department of Theoretical Physics, KLTE,
H-4010 Debrecen P.O.Box 5, Hungary}
\address[ATOMKI]{Institute of Nuclear Research of the Hungarian
Academy of Sciences, H-4001 Debrecen P.O.Box 51, Hungary}
\date{\today}

\begin{abstract}
We present the next-to-leading order partonic cross sections involving
an electroweak vector boson and four massless partons (quarks, gluons
or long living gluinos) in a general gauge theory with a simple Lie
Group. The vector boson couples to a massless lepton pair and a
quark-antiquark pair. The cross sections are given in terms of group
independent kinematical functions multiplying the eigenvalues of the
Casimir operators of the Lie group.  This kind of color decomposition
is required for the calculation of O($\alpha_s^3$) corrections to the
group independent kinematical functions in the four-jet production
cross sections in electron-positron annihilation. The knowledge of
these corrections facilitates the simultaneous precision meaurement of
the strong coupling and the color charge factors using the four-jet LEP
or SLC data as well as the test whether these data favour or exclude
the existence of a light gluino.

\end{abstract}
\end{frontmatter}


Electron-positron annihilation into four jets is a very clean way of
measuring the basic properties of Quantum Chromodynamics (QCD), the
theory of strong interactions. This process allows the simultaneous
measurement of the strong coupling $\as$ and the eigenvalues of the 
Casimir operators of the underlying symmetry group, the color factors.
Such a measurement provides a very stringent test of QCD because $\as$
is the only free parameter in the theory, while the color factors show
whether the dynamics is indeed described by an SU(3) symmetry.
However, in order to carry out such a full measurement the theoretical
prediction of perturbative QCD has to be known at \NLO accuracy. Lacking
this sort of precision of the theoretical description, so far the
experiments used four-jet data for measuring the ratios of the color
factors $C_A/C_F$ and $T_R/C_F$ that are relatively insensitive to
higher order corrections \cite{LEP}. The $\as$ measurements require the
fixing the normalization of four-jet observables, which however,
suffers large radiative corrections. For example, the experimental
four-jet rate measured as a function of the jet resolution parameter
was about a factor of two higher than the leading order prediction.

Recent theoretical developments make possible the \NLO calculation of 
four-jet quantities. There are now several general methods available
for the cancellation of infrared divergences that can be used for setting
up a Monte Carlo evaluation of \NLO partonic cross sections \cite{jets}.
The other vital piece of new information is the one-loop matrix elements
for the relevant QCD subprocesses --- i.e.\ for the production of
four-quarks and for the production of two quarks and two gluons in
electron-positron annihilation --- that are now available
\cite{GM4q,BDKW4q,CGM2q2g,BDK2q2g}. In refs.~\cite{GM4q,CGM2q2g}
Campbell, Glover and Miller make FORTRAN programs for the \NLO squared
matrix elements of the $e^+e^- \to \gamma^* \to \qb q \Qb Q$ and
$\qb q g g$ processes publicly available. In refs.~\cite{BDKW4q,BDK2q2g}
Bern, Dixon, Kosower and Wienzierl give analytic formul\ae\ for the
helicity amplitudes of the same processes with the $e^+e^- \to Z^0 \to$
four partons channel included as well. The helicity amplitudes for the
processes $e^+e^-\to \qb q ggg$ and $e^+e^-\to \qb q \Qb Q g$ have been
known for a long time \cite{a5parton}. Using the helicity amplitudes in
refs.~\cite{BDKW4q,BDK2q2g,a5parton}, Dixon and Signer calculated the \NLO
corrections to four-jet fractions for various clustering algorithms
\cite{DSjets}, as well as to angular distributions \cite{Signer} using
a general purpose partonic Monte Carlo program \cite{menloparc}. As a
result, the agreement between the data and the theoretical prediction
for the four-jet rates improved significantly.  In the calculation of
these four-jet fractions the underlying gauge group is assumed to be
SU($\Nc$) and the results are presented in the form of an expansion in
the number of colors. This way of presenting the results is sufficient
if one is interested in the four-jet rate and will certainly find
important application in LEP2 physics, where the $e^+e^-\to (\gamma, Z)
\to$ 4 jets events are the main background to $e^+e^-\to W^+W^- \to$ 4
jets events, but not for increasing the precision in measuring the
color charge factors.

One can measure the color charges directly using any kind of four-jet
observable, provided the partonic cross section is given in a factorized
form of eigenvalues of the Casimir operators and group indpendent
kinematical factors. Such a decomposition has been known for a long time 
at tree level \cite{ERT}. In this letter we give the necessary
decomposition of the virtual corrections. Combining these with
similarly decomposed real corrections, we can obtain the full 
O($\alpha_s^3$) corrections to the group independent kinematical
functions in the four-jet production cross sections in
electron-positron annihilation \cite{NTQCD97}.

An additional piece of motivation for using a group independent color
decomposition comes from the suggestion that the measurement of the color
charges may be influenced by the production of light, colored, but
electrically neutral fermionic particles, such as gluinos \cite{gluino}.
The effect of the presence of light gluinos on the strong
coupling is to increase $\as(M_Z)$ by 2\,\% \cite{ClavSur}. It would be
interesting to see if the presence of light gluinos influence the value
of the color charges more noticably. In order to make possible a \NLO
analysis of this effect, we also present the one-loop helicity
amplitudes for the $e^+e^-\to (\gamma, Z) \to \qb q \gt \gt$ process.
We have used these amplitudes to show that in order to obtain the most
significant light gluino exclusion limit from four-jet data, it is
advantageous to define jets in the range of small resolution parameter
\cite{NTgluino}. For example, in an analysis based upon ALEPH data for
four jet rates \cite{ALEPHR4} using Durham clustering algorithm
\cite{durham} at $y_{\rm cut}=0.002$ we were able to exclude the
existence of light gluinos at the 95\,\% confidence level
\cite{NTgluino}.

In presenting the new color decomposition, we rely on the work
of Bern, Dixon, Kosower and Wienzierl who published the explicit form
of the one-loop helicity amplitudes for the QCD subprocesses
\cite{BDKW4q,BDK2q2g}. We use the same notation as these articles and
introduce new ones but to the extent that is necessary.


Our aim is to give the \NLO squared matrix elements for the
$e^+e^-\to \qb q{\bar p}p$ processes ($p = Q$, $g$, or $\tilde{g}$)
in terms of color factors multiplied by group independent kinematic
functions. In order to find a group independent decomposition of the
squared matrix element, we have to give different color decompositions
of the one-loop helicity amplitudes for the various processes than those
presented in refs.~\cite{BDKW4q,BDK2q2g}, where the color charge
information has been lost by the use of the SU($\Nc$) Fierz identity and
SU($\Nc$) relations $C_F=(\Nc^2-1)/\Nc$, $C_A=2 \Nc$ (for $T_R=1$). In
the new color decomposition we can only use the defining relation of the
Lie algebra
\beq
\label{liealg}
[T^a,T^b]_{ij} = \i \sum_{c=1}^\NA f^{abc}T^c_{ij}\:,
\eeq
and the definition of the quadratic Casimir invariants $C_F$, $C_A$ and
$T_R$:
\beq
\label{quadcas}
\sum_{a=1}^\NA (T^aT^{\dag a})_{ij} = C_F \delta_{ij}\:,\quad
\sum_{a=1}^\NA (F^aF^{\dag a})_{cd} = C_A \delta_{cd}\:,\quad
\Tr(T^aT^{\dag b}) = T_R \delta^{ab}\:,
\eeq
where $\NA$ is the dimension of the adjoint representation of the gauge
group, $T^a$ are the generators in the fundamental and $F^a$ are those in
the adjoint representation. The latter are related to the structure
constans via $F^a_{bc} = -\i f_{abc}$. Careful analysis of the color
structure of the individual Feynman diagrams contributing to a given
process shows that the color charge information can completely be 
recovered from the primitive amplitudes of refs.~\cite{BDKW4q,BDK2q2g}
for the QCD subprocesses, while in the case of two-quark two-gluino
production minor modification of the partial amplitudes is necessary.
In order we can use those primitive amplitudes we use $T_R=1$
normalization.

Let us first consider the $e^+e^-\to \qb q\Qb Q$ process. At tree
level the new color decomposition of the
$\A60(1_q,2_\Qb,3_Q,4_\qb;5_\eb,6_e)$ helicity amplitude is
\beqn
\label{atree4q}
&&\A60(1_q,2_\Qb,3_Q,4_\qb;5_\eb,6_e)
= 2 e^2g^2 \sum_{c=1}^\NA T^c_{q\qb}T^c_{Q\Qb}
\\ \nn &&\qquad\qquad\times
\left[
 C_q^{h_qh_e}
\At_{6;0}(1^{h_q},2^{h_\Qb},3^{h_Q},4^{h_\qb};5^{h_\eb},6^{h_e})
\right.
\\ \nn &&\qquad\qquad\quad
\left.
+C_Q^{h_Qh_e}
\At_{6;0}(3^{h_Q},4^{h_\qb},1^{h_q},2^{h_\Qb};5^{h_\eb},6^{h_e})
\right],
\eeqn
where $e$ is the QED coupling, $g$ the QCD coupling $h_x$ is the helicity
of particle $x$ and the coupling factors $C_\cq^{h_\cq h_e}$ depend on the
charge $Q^\cq$ of quark $\cq$ in units of $e$ and on the left- and
right-handed couplings of the electron ($v_{L,R}^e$) and of the quark
($v_{L,R}^\cq$) as follows:
\beqn
&&C_\cq^{++}=-Q^\cq+v_R^\cq v_R^e \Ps56\:,
\qquad
C_\cq^{+-}=-Q^\cq+v_R^\cq v_L^e \Ps56\:,
\\ &&
C_\cq^{-+}=-Q^\cq+v_L^\cq v_R^e \Ps56\:,
\qquad
C_\cq^{--}=-Q^\cq+v_L^\cq v_L^e \Ps56\:.
\eeqn
The left- and right-handed couplings of the electron and quark are the
standard ones as given in ref.~\cite{BDKW4q}. $\At_{6;0}$ denotes the
tree-level partial amplitude $A_6^{\rm tree}$ of ref.~\cite{BDKW4q},
where it was defined to include a photon propagator. The ratio
$\Ps56 = s_{56}/(s_{56}-M_Z^2+\i \Gamma_Z M_Z)$
appearing in the coupling factors replaces this photon propagator with a
$Z$ propagator ($M_Z$ and $\Gamma_Z$ are the mass and width of the $Z^0$
boson).

At one loop the new color decomposition of the helicity amplitudes is
\beqn
\label{a1loop4q}
&&{\cal A}^{{\rm 1-loop}}_6(1_q,2_\Qb,3_Q,4_\qb;5_\eb,6_e)
= 2 e^2g^4
\\ \nn &&\qquad\times
\left\{
 C_q^{h_qh_e}
\left[
 \At_{6;1}(1^{h_q},2^{h_\Qb},3^{h_Q},4^{h_\qb};5^{h_\eb},6^{h_e})
 \sum_{c=1}^\NA T^c_{q\qb}T^c_{Q\Qb}
\right.
\right.
\\ \nn &&\qquad\qquad\qquad
\left.
+\At_{6;2}(1^{h_q},2^{h_\Qb},3^{h_Q},4^{h_\qb};5^{h_\eb},6^{h_e})
 \sum_{c,d=1}^\NA (T^cT^d)_{q\qb}(T^dT^c)_{Q\Qb}
\right]
\\ \nn &&\qquad\quad
+C_Q^{h_Qh_e}
\left[
 \At_{6;1}(3^{h_Q},4^{h_\qb},1^{h_q},2^{h_\Qb};5^{h_\eb},6^{h_e})
 \sum_{c=1}^\NA T^c_{q\qb}T^c_{Q\Qb}
\right.
\\ \nn &&\qquad\qquad\qquad
\left.
+\At_{6;2}(3^{h_Q},4^{h_\qb},1^{h_q},2^{h_\Qb};5^{h_\eb},6^{h_e})
 \sum_{c,d=1}^\NA (T^cT^d)_{q\qb}(T^dT^c)_{Q\Qb}
\right]
\\ \nn &&\qquad\quad
\left.
+C_{\rm ax}^{h_e}
 \At_{6;3}(1^{h_q},2^{h_\Qb},3^{h_Q},4^{h_\qb};5^{h_\eb},6^{h_e})
 \sum_{c=1}^\NA T^c_{q\qb}T^c_{Q\Qb}
\right\},
\eeqn
where $C_{\rm ax}^{h_e}$ vanishes for the photon and $W^\pm$ boson, while
for the $e^+e^-\to Z^0\to \qb q \Qb Q$ process it is
\beq
C_{\rm ax}^+ = \frac{v^e_R}{\sin 2\theta_W}\Ps56\:,\quad
C_{\rm ax}^- = \frac{v^e_L}{\sin 2\theta_W}\Ps56\:,
\eeq
with $\theta_W$ being the Weinberg angle.
In eq.~(\ref{a1loop4q}) we used the notation $\At_{6;i}$ for the partial
amplitudes in order to distinguish them from the $A_{6;i}$ functions
introduced in ref.~\cite{BDKW4q}, where the basic gauge invariant classes
of colorless amplitudes, the `primitive amplitudes' are also given
explicitly. Our new partial amplitudes can be expressed in terms of
the same primitive amplitudes multiplied by color factors. The expressions
depend on the helicities of the partons. There are only two independent
helicity configurations, which can be taken to be 
$(1_q^+,2_\Qb^\pm,3_Q^\mp,4_\qb^-;5_\eb^-,6_e^+)$, and from which the
amplitudes of the other helicity configurations can be obtained
\cite{BDKW4q}.

Formulas (\ref{atree4q}) and (\ref{a1loop4q}) apply to the case of
unequal quark flavors, $q\ne Q$. The equal flavor amplitude may be
obtained from the unequal-flavor formula by subtracting the same formula
with $q$ and $Q$ exchanged and then setting $Q=q$ in all the coupling
constant prefactors \cite{BDKW4q}.

The explicit expressions for the $\At_{6;i}$ partial amplitudes in
terms of primitive amplitudes are (we suppress the 5, 6 labels of the
lepton pair)
\beqn
\label{A614qpp}
&&
\At_{6;1}(1_q^+,2_\Qb^+,3_Q^-,4_\qb^-) =
\\ \nn &&\qquad
-C_F A_6^{\rm sl}(2,3,1,4)
+\frac{C_A}{2}\left(A_6^{\rm sl}(2,3,1,4)-A_6^{+-}(1,3,2,4)\right)
\\ \nn &&\qquad
+A_6^{t,++}(1,2,3,4)
     - \Nf \left(A_6^{f,++}(1,2,3,4)+A_6^{s,++}(1,2,3,4)\right)\:,
\\ &&
\At_{6;2}(1_q^+,2_\Qb^+,3_Q^-,4_\qb^-) =
A_6^{++}(1,2,3,4)+A_6^{+-}(1,3,2,4)\:,
\\ &&
\At_{6;3}(1_q^+,2_\Qb^+,3_Q^-,4_\qb^-) = A_6^{\rm ax}(1,4,2,3)\:,
\eeqn
while for the other helicity configuration
\beqn
\label{A614qpm}
&&
\At_{6;1}(1_q^+,2_\Qb^-,3_Q^+,4_\qb^-) =
\\ \nn &&\qquad
 C_F A_6^{\rm sl}(3,2,1,4)
-\frac{C_A}{2}\left(A_6^{\rm sl}(3,2,1,4)+A_6^{++}(1,3,2,4)\right)
\\ \nn &&\qquad
+A_6^{t,+-}(1,2,3,4)
     - \Nf \left(A_6^{f,+-}(1,2,3,4)+A_6^{s,+-}(1,2,3,4)\right)\:,
\\ &&
\At_{6;2}(1_q^+,2_\Qb^-,3_Q^+,4_\qb^-) =
A_6^{++}(1,3,2,4)+A_6^{+-}(1,2,3,4)\:,
\\ &&
\At_{6;3}(1_q^+,2_\Qb^-,3_Q^+,4_\qb^-) = -A_6^{\rm ax}(1,4,3,2)\:.
\eeqn

The gluino is a majorana fermion in the adjoint representation of the
gauge group and does not couple to the vector bosons directly. Therefore,
the $e^+e^-\to \qb q \gt \gt$ subprocesses has similar color decomposition
to that of the $e^+e^-\to \qb q \Qb Q$ subprocess.  At tree level this
decomposition reads
\beqn
\label{atree2q2gt}
&&\A60(1_q,2_\gt,3_\gt,4_\qb;5_\eb,6_e) =
\\ \nn &&\qquad
2 e^2g^2 C_q^{h_qh_e}
 \At_{6;0}(1^{h_q},2^{h_2},3^{h_3},4^{h_\qb};5^{h_\eb},6^{h_e})
 \sum_{c=1}^\NA T^c_{q\qb}F^c_{\gt_2\gt_3}\:,
\eeqn
while at one-loop it is
\beqn
\label{a1loop2q2gt}
&&{\cal A}^{{\rm 1-loop}}_6(1_q,2_\gt,3_\gt,4_\qb;5_\eb,6_e)
= 2 e^2g^4
\\ \nn &&\qquad\times
\left\{
C_q^{h_qh_e}
\left[
 \At_{6;1}(1^{h_q},2^{h_2},3^{h_3},4^{h_\qb};5^{h_\eb},6^{h_e})
 \sum_{c=1}^\NA T^c_{q\qb}F^c_{\gt_3\gt_2}
\right.
\right.
\\ \nn &&\qquad\qquad\qquad
\left.
+\At_{6;2}(1^{h_q},2^{h_2},3^{h_3},4^{h_\qb};5^{h_\eb},6^{h_e})
 \sum_{c,d=1}^\NA (T^cT^d)_{q\qb}(F^dF^c)_{\gt_3\gt_2}
\right]
\\ \nn &&\qquad\quad\;
+\:C_{\rm ax}^{h_e}
\left.
\left[
 \At_{6;3}(1^{h_q},2^{h_2},3^{h_3},4^{h_\qb};5^{h_\eb},6^{h_e})
 \sum_{c=1}^\NA T^c_{q\qb}F^c_{\gt_3\gt_2}
\right]
\right\}.
\eeqn

The $\At_{6;i}$ partial amplitudes for the $e^+e^-\to \qb q \gt \gt$
process are closely related to those of the $e^+e^-\to \qb q \Qb Q$
process. In fact, the $\At_{6;2}$ and $\At_{6;3}$ amplitudes are
exactly the same,
\beqn
&&
\At_{6;2}(1_q^+,2_\gt^\pm,3_\gt^\mp,4_\qb^-) =
\At_{6;2}(1_q^+,2_\Qb^\pm,3_Q^\mp,4_\qb^-)\:,
\\ &&
\At_{6;3}(1_q^+,2_\gt^\pm,3_\gt^\mp,4_\qb^-) =
\At_{6;3}(1_q^+,2_\Qb^\pm,3_Q^\mp,4_\qb^-)\:,
\eeqn
while the $\At_{6;1}$ amplitudes differ in terms arising from fermion
bubble and parent triangle diagrams \cite{BDKW4q},
\beqn
&&
\At_{6;1}(1_q^+,2_\gt^\pm,3_\gt^\mp,4_\qb^-) =
\At_{6;1}(1_q^+,2_\Qb^\pm,3_Q^\mp,4_\qb^-)
-(C_F - C_A)A_6^{\Delta,+\pm}(1,2,3,4)
\nn \\ &&\qquad\qquad\qquad\qquad\quad\;\;
-N_\gt \frac{C_A}{2} \left(A_6^{f,+\pm}(1,2,3,4)+A_6^{s,+\pm}(1,2,3,4)\right),
\eeqn
where $N_\gt$ is the number of light gluino flavors and 
\beqn
&&A_6^{\Delta,+\pm}(1,2,3,4)=
\nn \\ &&\qquad
c_\Gamma A_6^{{\rm tree,}+\pm}(1,2,3,4)
\left[
-\frac{1}{\ep^2}\left(\frac{\mu^2}{-s_{23}}\right)^\ep
-\frac{3}{2\ep}\left(\frac{\mu^2}{-s_{23}}\right)^\ep-\frac{7}{2}
\right],
\eeqn
with $c_\Gamma$ and $A_6^{{\rm tree,}++}$ given in ref.~\cite{BDKW4q}. We
remark here that gluinos are Majorana fermions therefore, the cross
section requires an identical-particle factor of $\frac{1}{2}$ in the
phase-space measure.

Finally, the new color decomposition of the helicity amplitudes for the
$e^+e^-\to \qb q g g$ process at tree level is given by
\beqn
\label{atree2q2g}
&&{\cal A}^{{\rm tree}}_6(1_q,2_g,3_g,4_\qb;5_\eb,6_e)
= 2 e^2g^2
\\ \nn &&\qquad\quad\times
C_q^{h_qh_e}
\left[
 \At_{6;0}(1^{h_q},2^{h_2},3^{h_3},4^{h_\qb};5^{h_\eb},6^{h_e})
 (T^{g_2}T^{g_3})_{q\qb}
\right.
\\ \nn &&\qquad\qquad\qquad
\left.
+\At_{6;0}(1^{h_q},3^{h_3},2^{h_2},4^{h_\qb};5^{h_\eb},6^{h_e})
 (T^{g_3}T^{g_2})_{q\qb}
\right],
\eeqn
and at one-loop it is
\beqn
\label{a1loop2q2g}
&&{\cal A}^{{\rm 1-loop}}_6(1_q,2_g,3_g,4_\qb;5_\eb,6_e)
= 2 e^2g^4
\\ \nn &&\qquad\times
\left\{
C_q^{h_qh_e}
\left[
 \At_{6;1}(1^{h_q},2^{h_2},3^{h_3},4^{h_\qb};5^{h_\eb},6^{h_e})
 (T^{g_2}T^{g_3})_{q\qb}
\right.
\right.
\\ \nn &&\qquad\qquad\qquad
+\At_{6;2}(1^{h_q},2^{h_2},3^{h_3},4^{h_\qb};5^{h_\eb},6^{h_e})
 (T^{g_3}T^{g_2})_{q\qb}
\\ \nn &&\qquad\qquad\qquad
\left.
+\At_{6;3}(1^{h_q},2^{h_2},3^{h_3},4^{h_\qb};5^{h_\eb},6^{h_e})
 \sum_{c,d=1}^\NA (T^cT^d)_{q\qb}(F^dF^c)_{g_3g_2}
\right]
\\ \nn &&\qquad\quad\;
+\:\sum_{f=1}^\Nf \frac{1}{2}\left(C_{q_f}^{+h_e} + C_{q_f}^{-h_e}\right)
\At_{6;4}(1^{h_q},2^{h_2},3^{h_3},4^{h_\qb};5^{h_\eb},6^{h_e})
\\ \nn &&\qquad\qquad\qquad\quad\;
\times \sum_{c=1}^\NA
\left[\Tr (T^{g_2}T^{g_3}T^c) T^c_{q\qb}
    + \Tr (T^{g_3}T^{g_2}T^c) T^c_{q\qb}\right]
\\ \nn &&\qquad\quad\;
+ C_{\rm ax}^{h_e}
\Bigg[
 \At_{6;5}^A(1^{h_q},2^{h_2},3^{h_3},4^{h_\qb};5^{h_\eb},6^{h_e})
 (T^{g_2}T^{g_3})_{q\qb}
\\ \nn &&\qquad\qquad\qquad
+\At_{6;5}^B(1^{h_q},2^{h_2},3^{h_3},4^{h_\qb};5^{h_\eb},6^{h_e})
 (T^{g_3}T^{g_2})_{q\qb}
\\ \nn &&\qquad\qquad\qquad
+ \At_{6;5}^C(1^{h_q},2^{h_2},3^{h_3},4^{h_\qb};5^{h_\eb},6^{h_e})
\\ \nn &&\qquad\qquad\qquad\;
\left.
\times \sum_{c=1}^\NA
\left[\Tr (T^{g_2}T^{g_3}T^c) T^c_{q\qb}
    + \Tr (T^{g_3}T^{g_2}T^c) T^c_{q\qb}\right]
\Bigg]
\right\}.
\eeqn

The partial amplitudes $\At_{6;i}$ can easily be constructed from the
primitive amplitudes of ref.~\cite{BDK2q2g}:
\beqn
\label{A612q2g}
&&
\At_{6;1}(1_q^+,2_g^{h_2},3_g^{h_3},4_\qb^-) =
C_F A_6(1_q,4_\qb;3,2)
\\ \nn &&\qquad
+\frac{C_A}{2}
\left[A_6(1_q,2,3,4_\qb)-A_6(1_q,4_\qb;3,2)-A_{6;3}(1_q,4_\qb;3,2)\right]
\\ \nn &&\qquad
+[A_6^{t,++}(1_q,2,3,4_\qb)
  - \Nf \left(A_6^{f,++}(1_q,2,3,4_\qb)
             +A_6^{s,++}(1_q,2,3,4_\qb)\right)\:,
\\ &&
\label{A622q2g}
\At_{6;2}(1_q^+,2_g^{h_2},3_g^{h_3},4_\qb^-) =
C_F A_6(1_q,4_\qb;2,3)
\\ \nn &&\qquad
+\frac{C_A}{2}\left[A_6(1_q,3,2,4_\qb)-A_6(1_q,4_\qb;2,3)\right]
\\ \nn &&\qquad
+A_6^{t,++}(1_q,3,2,4_\qb)
  - \Nf \left(A_6^{f,++}(1_q,3,2,4_\qb)
             +A_6^{s,++}(1_q,3,2,4_\qb)\right)\:,
\\ &&
\At_{6;3}(1_q^+,2_g^{h_2},3_g^{h_3},4_\qb^-)
= A_{6;3}(1_q^+,4_\qb^-;2_g^{h_2},3_g^{h_3})\:,
\\ &&
\At_{6;4}(1_q^+,2_g^{h_2},3_g^{h_3},4_\qb^-)
= A_{6;4}^{{\rm v}}(1_q^+,4_\qb^-;2_g^{h_2},3_g^{h_3})\:,
\\ &&
\At_{6;5}^A(1_q^+,2_g^{h_2},3_g^{h_3},4_\qb^-)
=\frac{1}{2}
\left[A_{6;4}^{\rm ax}(1_q^+,4_\qb^-;2_g^{h_2},3_g^{h_3})
    - A_{6;4}^{\rm ax}(1_q^+,4_\qb^-;3_g^{h_3},2_g^{h_2})
\right.
\nn \\ && \qquad\qquad\qquad\qquad\qquad\quad
\left.
    + A_{6;5}^{\rm ax}(1_q^+,4_\qb^-;2_g^{h_2},3_g^{h_3})\right]\:,
\\ &&
\At_{6;5}^B(1_q^+,2_g^{h_2},3_g^{h_3},4_\qb^-)
=\frac{1}{2}
\left[A_{6;4}^{\rm ax}(1_q^+,4_\qb^-;3_g^{h_3},2_g^{h_2})
    - A_{6;4}^{\rm ax}(1_q^+,4_\qb^-;2_g^{h_2},3_g^{h_3})
\right.
\nn \\ && \qquad\qquad\qquad\qquad\qquad\quad
\left.
    + A_{6;5}^{\rm ax}(1_q^+,4_\qb^-;2_g^{h_2},3_g^{h_3})\right]\:,
\\ &&
\At_{6;5}^C(1_q^+,2_g^{h_2},3_g^{h_3},4_\qb^-)
=\frac{1}{2}
\left[A_{6;4}^{\rm ax}(1_q^+,4_\qb^-;2_g^{h_2},3_g^{h_3})
    + A_{6;4}^{\rm ax}(1_q^+,4_\qb^-;3_g^{h_3},2_g^{h_2})
\right.
\nn \\ && \qquad\qquad\qquad\qquad\qquad\quad
\left.
    - A_{6;5}^{\rm ax}(1_q^+,4_\qb^-;2_g^{h_2},3_g^{h_3})\right]\:.
\eeqn

Before closing the discussion of the helicity amplitudes, we should
remark that in order to take into account the effect of light gluinos in
a \NLO calculation fully, one has to make the change
\beq
\Nf \to \Nf + \frac{C_A}{2} N_\gt
\eeq
in eqs.~(\ref{A614qpp}), (\ref{A614qpm}) and (\ref{A612q2g}),
(\ref{A622q2g}), where in SU($\Nc$) theory $C_A=2\Nc$ for $T_R=1$.


As a next step we use eqs.~(\ref{atree4q}), (\ref{a1loop4q}),
(\ref{atree2q2gt}), (\ref{a1loop2q2gt}), and (\ref{atree2q2g}),
(\ref{a1loop2q2g}) to give a gauge independent decomposition of the
\NLO squared matrix elements. The general expression for these matrix
elements is
\beq
\label{dsigmanlo}
\d \sigma_6^{{\rm NLO,\,virtual}}
= 2\sum_{{\rm helicities}} \sum_{{\rm colors}} 
{\rm Re}\,[{\cal A}^{{\rm tree}\:*}_6 {\cal A}^{{\rm 1-loop}}_6]\:.
\eeq
We want to evaluate the color sum in such a way that the group
independent information is maintained.  For this purpose, we use the
commutation relation (\ref{liealg}) together with the definition of the
quadratic Casimirs, eq.~(\ref{quadcas}) to derive the necessary
Lie-algebra relations.  In the case of production of two unequal flavor
quark pairs these are:
\beqn
\label{trTTtrTT}
&&\quad
\sum_{a,b=1}^\NA \Tr(T^a T^{\dag b}) \Tr(T^a T^{\dag b}) = \Nc C_F\:,
\\ &&
\label{cubiccas}
\sum_{a,b,c=1}^\NA \Tr(T^a T^b T^{\dag c}) \Tr(T^{\dag c} T^b T^a)
\equiv C_3\:,
\eeqn
where in eq.~(\ref{cubiccas}) we introduced $C_3$, the square of a
cubic Casimir, that cannot be reduced to an expression of the quadratic
Casimirs $C_F$ and $C_A$. In SU($\Nc$) $C_3$ takes the value
$(\Nc^2-1)(\Nc^2-2)/\Nc$.
In the case of production of two equal flavor quarks, in addition to
relations (\ref{trTTtrTT}) and (\ref{cubiccas}), we need two more color
sums:
\beqn
\label{trTTTT}
&&\qquad
\sum_{a,b=1}^\NA \Tr(T^a T^{\dag b} T^a T^{\dag b})
= \Nc C_F\left(C_F-\frac{C_A}{2}\right)\:,
\\ &&
\label{trTTTTTT}
\sum_{a,b,c=1}^\NA \Tr(T^a T^b T^{\dag c} T^b T^a T^{\dag c})
= \Nc C_F\left(C_F-\frac{C_A}{2}\right)^2.
\eeqn
When a quark and a gluino pair appears in the final state, we use
relations
\beqn
\label{trTTtrFF}
&&\qquad
\sum_{a,b=1}^\NA \Tr(T^a T^{\dag b}) \Tr(F^a F^{\dag b}) = \Nc C_F C_A\:,
\\ &&
\label{trTTTtrFFF}
\sum_{a,b,c=1}^\NA \Tr(T^a T^b T^{\dag c}) \Tr(F^{\dag c} F^b F^a)
= \Nc C_F \left(\frac{C_A}{2}\right)^2,
\eeqn
Finally, for the production of a quark pair and two gluons the additional
necessary color sums are
\beqn
\label{trTTTT2}
&&\qquad\qquad
\sum_{a,b=1}^\NA \Tr(T^a T^b T^{\dag b} T^{\dag a}) = \Nc C_F^2\:,
\\ &&
\label{trTTTTFF}
\sum_{a,b,g_2,g_3=1}^\NA \Tr(T^a T^b T^{\dag g_3} T^{\dag g_2})
(F^b F^a)_{g_3g_2} = \Nc C_F\left(\frac{C_A}{2}\right)^2,
\\ &&\qquad
\label{trTTTTFF2}
\sum_{a,b,g_2,g_3=1}^\NA \Tr(T^a T^b T^{\dag g_2} T^{\dag g_3})
(F^b F^a)_{g_3g_2} = 0\:,
\eeqn
and
\beq
\label{trTTTtrTTT}
\sum_{g_2,g_3,c=1}^\NA
\Tr(T^{g_2} T^{g_3} T^c) \Tr(T^{\dag g_2} T^{\dag g_3} T^c)
=C_3-C_F\frac{C_A}{2}\Nc\:.
\eeq

Using eqs.~(\ref{trTTtrTT}-\ref{trTTTtrTTT}), one can evaluate the
color sum in eq.~(\ref{dsigmanlo}) easily. The resulting
differential cross sections can be written as a quadratic form:
\beqn
&&\frac{1}{\sigma_0}\d \sigma_6^{{\rm NLO,\,virtual}}(\vec{p})
= \left(\frac{\as C_F}{2\pi}\right)^3
\\ \nn &&\qquad\qquad\qquad\times
\left[
C_0(\vec{p}) + C_x(\vec{p})\,x + C_y(\vec{p})\,y + C_z(\vec{p})\,z
\right.
\\ \nn &&\quad\qquad\qquad\qquad
\left.
+ C_{xx}(\vec{p})\,x^2 + C_{xy}(\vec{p})\,x\,y + C_{yy}(\vec{p})\,y^2
\right]\:.
\eeqn
In this equation $\sigma_0$ denotes the Born cross section for the
process $e^+e^-\to \qb q$, $\vec{p}$ is the collection of the final state
momenta, and $x$, $y$ and $z$ are ratios of eigenvalues of the
Casimir operators defined as
\beq
x=\frac{C_A}{C_F}\:,\quad
y=\frac{T_R}{C_F}=\frac{\Nc}{\NA}\:,\quad
z=\frac{C_3}{\Nc C_F^2}\:.
\eeq
These ratios are the sole quantities together with the overall
normalization that carry group information. The coefficients $C_i(\vec{p})$
are the group independent kinematical functions that depend on the
subprocess. Their explicit expressions are quite complicated, but can be
obtained straightforwardly using the formulas of this letter. These
expressions will be published in the form of C++ code in the partonic
Monte Carlo program DEBRECEN.


In this letter we have shown explicitly how to rewrite the one-loop
helicity amplitudes of the $e^+e^-\to (\gamma, Z) \to$ 4 partons
processes in a form from which the group independent color
decomposition of the \NLO squared matrix elements for these processes
can be obtained. These kinematical functions for the QCD subprocesses
can completely be written in terms of the primitive amplitudes given in
refs.~\cite{BDKW4q} and \cite{BDK2q2g}. In the case of the $e^+e^-\to \qb
q \gt \gt$ subprocess one has to modify the primitive amplitudes of the
four-quark subprocess slightly. We also presented the general structure of
the \NLO partonic cross sections in terms of group independent kinematical
functions multiplying ratios of eigenvalues of the Casimir operators of
the Lie group. This kind of color decomposition together with a similar
decomposition of the $e^+e^-\to (\gamma, Z) \to$ 5  partons squared
matrix elements is required for the calculation of O($\alpha_s^3$)
corrections to the group independent kinematical functions in the
four-jet production cross sections in electron-positron annihilation.
We anticipate that these corrections will improve our knowledge of the
basic parameters of QCD.

\bigskip
This research was supported in part by the EEC Programme "Human Capital
and Mobility", Network "Physics at High Energy Colliders", contract
PECO ERBCIPDCT 94 0613 as well as by the Hungarian Scientific Research
Fund grant OTKA T-016613 and the Research Group in Physics of the
Hungarian Academy of Sciences, Debrecen.

\def\np#1#2#3  {Nucl.\ Phys.\ {\bf #1}, #2 (19#3)}
\def\pl#1#2#3  {Phys.\ Lett.\ {\bf #1}, #2 (19#3)}
\def\prep#1#2#3  {Phys.\ Rep.\ {\bf #1}, #2 (19#3)}
\def\prd#1#2#3 {Phys.\ Rev.\ D {\bf #1}, #2 (19#3)}
\def\prl#1#2#3 {Phys.\ Rev.\ Lett.\ {\bf #1}, #2 (19#3)}
\def\zpc#1#2#3  {Zeit.\ Phys.\ C {\bf #1}, #2 (19#3)}
\def\cmc#1#2#3  {Comp.\ Phys.\ Comm.\ {\bf #1}, #2 (19#3)}


\begin{thebibliography}{99}
\bibitem{LEP}
B. Adeva et al, L3 Collaboration, \pl{B248}{227}{97} ;\\
P. Abreu et al, DELPHI Collaboration, \zpc{59}{357}{93} ;\\
R. Akers et al, OPAL Collaboration, \zpc{65}{367}{95} ;\\
R. Barate et al, ALEPH Collaboration, preprint CERN-PPE/97-002.
\bibitem{jets}
W.T. Giele and E.W.N. Glover, \prd{46}{1980}{92} ;\\
S. Frixione, Z. Kunszt and A. Signer, \np{B467}{399}{96} ;\\
S. Catani and M.H. Seymour, \pl{B378}{287}{96} ; \np{485}{291}{97} ;\\
Z. Nagy and Z. Tr\'ocs\'anyi, \np{486}{189}{97} .
\bibitem{GM4q} E.W.N. Glover and D.J. Miller, \pl{B396}{257}{97} .
\bibitem{BDKW4q} Z. Bern, L. Dixon, D. A. Kosower and S. Wienzierl,
\np{B489}{3}{97} ;
\bibitem{CGM2q2g} J.M. Campbell, E.W.N. Glover and D.J. Miller, 
preprint hep-ph/9706297.
\bibitem{BDK2q2g} Z. Bern, L. Dixon and D. A. Kosower,
preprint hep-ph/9708239.
\bibitem{a5parton}
K. Hagiwara and D. Zeppenfeld, \np{B313}{560}{89} ;\\
F.A. Berends, W.T. Giele and H. Kuijf, \np{B321}{39}{89} ;\\
N.K. Falk, D. Graudenz and G. Kramer, \np{B328}{317}{89} .
\bibitem{DSjets}
A. Signer and L.Dixon, \prl{78}{811}{97} ;\\
L.Dixon and A. Signer, preprint hep-ph/9706285.
\bibitem{Signer} A. Signer, preprint hep-ph/9705218.
\bibitem{menloparc}
A. Signer, preprint SLAC-PUB-7531 (1997).
\bibitem{ERT} R.K. Ellis, D.A. Ross and A.E. Terrano, \np{B178}{421}{81} .
\bibitem{NTQCD97} Z. Nagy and Z. Tr\'ocs\'anyi, preprint hep-ph/9708344.
\bibitem{gluino}
G.R. Farrar, \pl{B265}{395}{91} ; \prd{51}{3904}{95} ; preprint
hep-ph/9707467;\\
J. Ellis, D. Nanopoulos and D. Ross, \pl{B305}{375}{93} ;\\
R. Mu$\tilde{\rm n}$oz-Tapia and W.J. Stirling, \prd{49}{3763}{94} ;\\
S. Moretti, R. Mu$\tilde{\rm n}$oz-Tapia and K. Odagiri, \pl{B389}{545}{96} ;
preprint hep-ph/9609295; \\
A. de Gouvea and H. Murayama, preprint hep-ph/9606449.
\bibitem{ClavSur} L.J. Clavelli and L.R. Surguladze, \prl{78}{1632}{97} .
\bibitem{NTgluino} Z. Nagy and Z. Tr\'ocs\'anyi, preprint hep-ph/9708343.
\bibitem{ALEPHR4} R. Barate et al, ALEPH collaboration, preprint
CERN-PPE-96-186.
\bibitem{durham} S. Catani Yu.L. Dokshitzer, M. Olsson, G. Turnock and
B.R. Webber, \pl{B269}{432}{91} .
\end{thebibliography}
\end{document}